\documentclass[12pt,preprint]{aastex}
\usepackage{psfig}
\usepackage{apjfonts}
\usepackage{mathrsfs}

\begin{document}

\title{The helium abundance in the metal-poor globular clusters M30 and NGC6397\footnote{Based on data 
taken at the ESO, within the observing programs 081.D-0356 and 087.D-0748.}}

\author{A. Mucciarelli$^{1}$, L. Lovisi$^{1}$, B. Lanzoni$^{1}$, F. R. Ferraro$^{1}$  }

\affil{$^{1}$Dipartimento di Fisica \& Astronomia, Universit\`a 
degli Studi di Bologna, Viale Berti Pichat, 6/2 - 40127
Bologna, ITALY}

\begin{abstract}

We present the helium abundance of the two metal-poor clusters M30 and NGC6397. Helium estimates have been 
obtained by using the high-resolution spectrograph FLAMES at the ESO Very Large Telescope and by measuring 
the He~I line at 4471 $\mathring{A}$ in 24 and 35 horizontal branch stars in M30 and NGC6397, respectively. 
This sample represents the largest dataset of He abundances collected so far in metal-poor clusters. 
The He mass fraction turns out to be Y=0.252$\pm$0.003 ($\sigma$=0.021) for M30 and 
Y=0.241$\pm$0.004 ($\sigma$=0.023) NGC6397. These values are fully 
compatible with the cosmological abundance, thus suggesting that the horizontal branch stars 
are not strongly enriched in He. The small spread of the Y distributions are compatible with those 
expected from the observed main sequence splitting. Finally, we find an hint of a weak anticorrelation 
between Y and [O/Fe] in NGC6397 in agreement with the prediction that O-poor stars are formed 
by (He-enriched) gas polluted by the products of hot proton-capture reactions.

\end{abstract}  
 

\section{Introduction}   
\label{intro}

Helium is the most abundant among the few chemical elements 
($^{3}$He, $^{4}$He, D, $^6$Li, $^7$Li, $^9$Be, $^{10}$B and $^{11}$B) 
synthesized directly in the primordial {\sl furnax} 
of the Big Bang. The most recent determination of the primordial He mass fraction
provides an initial value $Y_{P}$=~0.254$\pm$0.003 \citep{izotov}.

The study of the He content of stars in globular clusters (GCs) 
is still a challenging task but it is crucial for a number of aspects 
of the stellar astrophysics.
First of all, the He content in Galactic GC stars is thought to be a good tracer of the 
primordial He abundance because these are among the first generations of stars formed in the Universe and 
the mixing episodes occurring during their evolution 
only marginally affect their surface He abundance \citep{sweigart97}. 
Moreover, the He content is usually invoked as one of the possible {\it second parameter} 
\citep[together with age, CNO/Fe ratio, stellar density; see e.g.][]{gratton10,dotter10,dalex13,milone13}, 
 to explain the observed distribution of stars along the horizontal branch (HB),
 being the overall metallicity the first parameter.
Finally, observational evidence reveal the presence of multiple stellar generations
in GCs, formed in  short timescales ($\sim$100 Myr) after the initial star-formation burst, 
from a pristine 
gas polluted by the products of hot proton-capture processes 
\citep[see e.g.][and references therein]{gratton12a}. 
Thus, these new stars are expected to be characterized by (mild or extreme) He enhancement 
with respect to the first ones, 
together with enhancement of Na and Al, and depletion of O and Mg.

Despite such an importance, however, the intrinsic difficulties in the derivation of He abundances 
in low-mass stars have prevented a detailed and systematic investigation of He in GCs. 
Only few  photospheric He transitions are available in the blue-optical
spectral range ($<$5900 $\mathring{A}$) and they are visible only at high effective temperatures ($T_{eff}$). 
Therefore, He lines in GC stars can be detected only among the HB stars
hotter than $\sim$9000 K (the precise boundary also depends on the available 
signal-to-noise ratio of the spectra, SNR).

Instead, the measure of the He abundance in FGK-type stars is limited only to the 
use of the chromospheric line at 10830 $\mathring{A}$, while no photospheric He line 
is available in these stars. Unfortunately, this transition is extremely weak and very high SNR 
and spectral resolution are required for a proper measurement. Moreover, the precise He abundance 
heavily depends on the modeling of the chromosphere. However, this line can provide differential 
measures of the He abundance, as performed by \citet{pasquini11} in two giants in NGC2808, 
\citet{dupree11} in 12 giant in Omega Centauri and \citet{dupree13} in two giants 
in Omega Centauri. \citet{pasquini11} point out a Y difference of at least 0.17 between the two stars. 
A similar difference has been suggested by \citet{dupree13} for giants in Omega Centauri.

A further complication in the measurement of the He abundance in HB stars 
is provided by diffusion processes, like radiative levitation and 
gravitational settling, occurring in the radiative atmospheres of HB stars 
hotter than $\sim$11000-12000 K, corresponding to the so-called {\it Grundahl Jump} 
\citep{grundahl99}.
These phenomena lead to a substantial modification of the surface chemical composition, 
and in particular to a decrease of the He abundance \citep[see Fig.~22 in][]{behr03} and an 
enhancement of the iron-peak element abundances. 
As a consequence, only HB stars in the narrow $T_{eff}$ range between $\sim$9000 and $\sim$11000 K 
can be used as reliable diagnostics of the He content of the parent cluster.

At present, determinations of the He mass fraction (Y) in GC HB stars not 
affected by diffusion processes
have been obtained only for some metal-intermediate ([Fe/H]$\sim$--1.5/--1.1) GCs: 
NGC6752 \citep[][$<$Y$>$=~0.24$\pm$0.01, 4 stars]{villanova09}, 
M4 \citep[][$<$Y$>$=~0.29$\pm$0.01, 6 stars]{villanova12}, 
NGC1851 \citep[][$<$Y$>$=~0.29$\pm$0.05, 20 stars]{gratton12b}, 
M5 \citep[][$<$Y$>$=~0.22$\pm$0.03, 17 stars]{gratton13}, 
NGC2808 \citep[][$<$Y$>$=~0.34$\pm$0.01, 17 stars]{marino13} 
and M22 \citep[][$<$Y$>$=~0.34$\pm$0.01, 29 stars]{gratton14}.
All these analyses are based on the photospheric He~I line at 5875 $\mathring{A}$.

Some evidence suggest that the variation of He in GC stars is linked to 
different chemical compositions. The differential analysis performed by \citet{pasquini11} 
on two giants in NGC2808 with different Na content highlights that the Na-rich star 
is also He enriched at odds with the Na-poor one. 
\citet{villanova09} and \citet{villanova12} derived He, Na and O abundances for HB stars in 
NGC6752 and M4, respectively, finding that the stars along the reddest part of the HB of NGC6752 
have a standard He content, as well as Na and O abundances compatible with the first generation, 
while the stars in the bluest part of the HB of M4 are slightly He-enhanced (by $\sim$0.05), 
with Na and O abundance ratios compatible with the second stellar generation. 
In a similar way, \citet{marino13} found a clear evidence of He enhancement (by $\sim$0.09) 
among the bluest HB stars in NGC2808, that are also all Na-rich.

Further spectroscopic evidence (not including the measure of He abundances) strengthen the connection 
between the HB morphology and the chemical composition, pointing out that the bluest portion of the 
HB (before the onset of the radiative levitation) is populated mainly by second generation stars, while 
the reddest part of the sequence is dominated by first generation stars \citep[like in M4,][]{marino11}
or by a mixture of first and second generation stars \citep[like in NGC2808,][]{marino13}.

In this paper we present the first determination of the He abundance
in HB stars of the metal-poor GCs M30 and NGC6397 
([Fe/H]=~--2.28$\pm$0.01 and [Fe/H]=~--2.12$\pm$0.01, \citet{lovisi12} and \citet{lovisi13}, 
respectively).

\section{Observations}
\label{obs}

\begin{figure*}
\plottwo{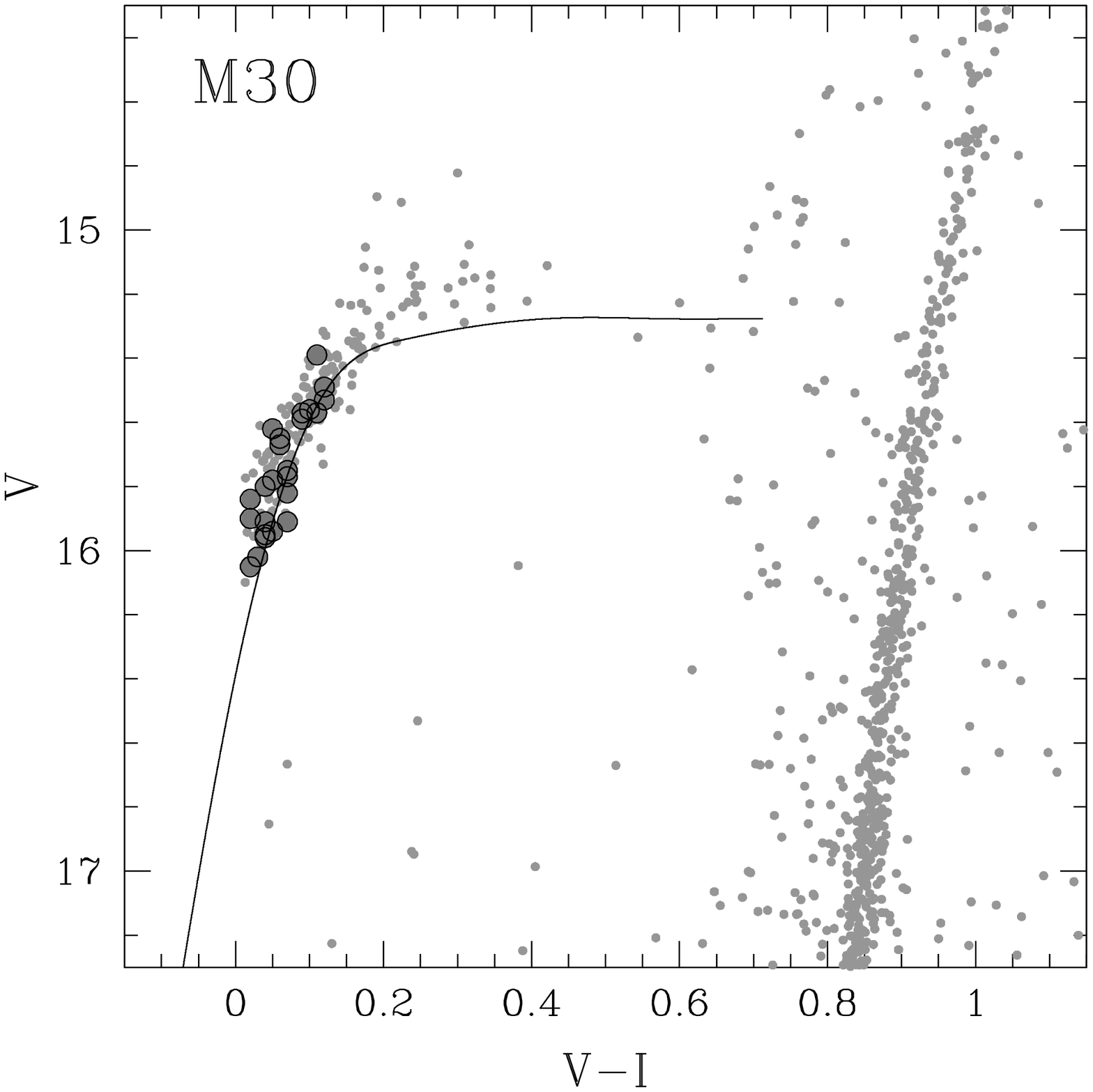}{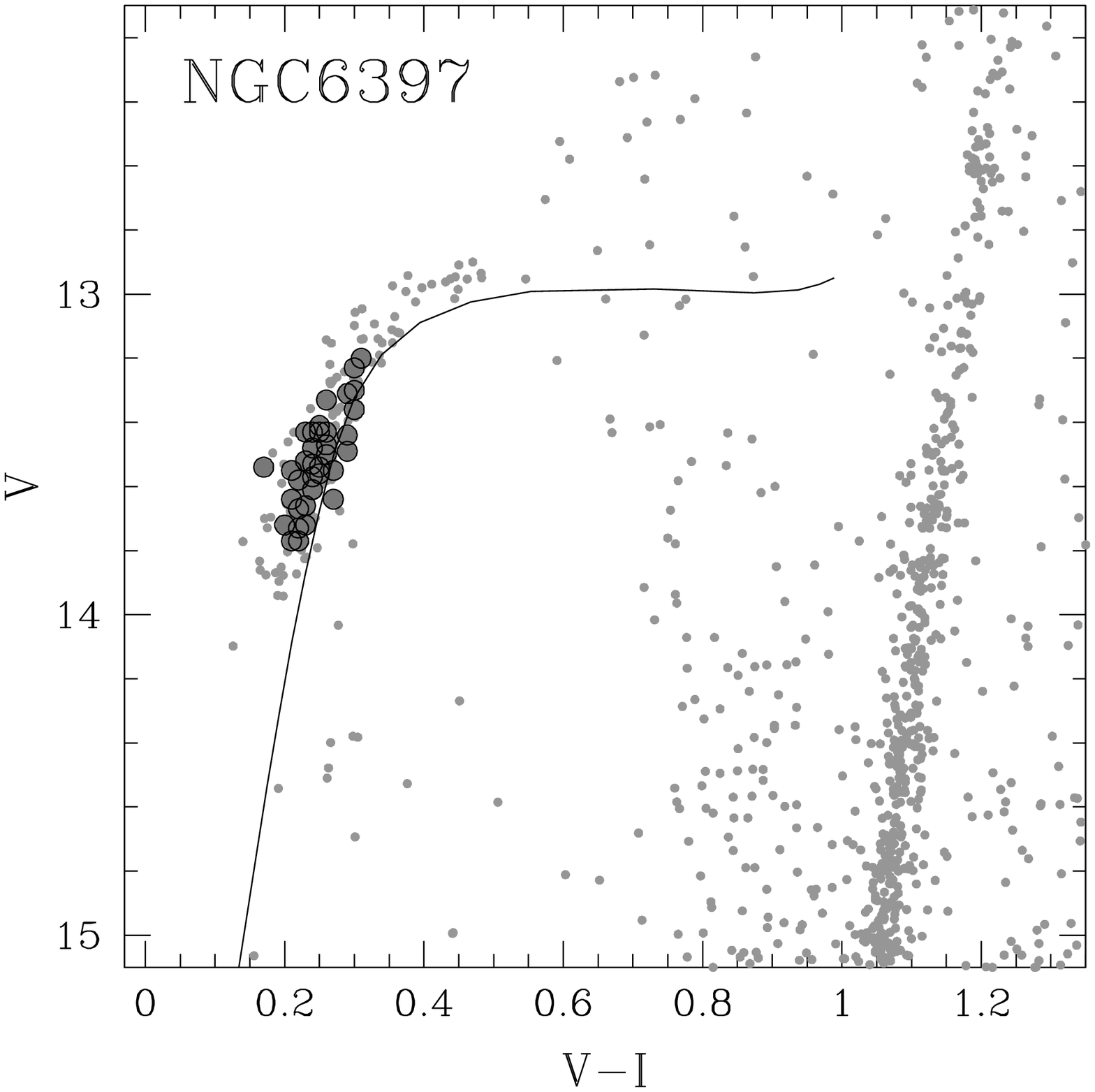}
\caption{
Color-magnitude diagram of M30 \citep[][left panel]{ferraro09a} 
and of NGC6397 (Contreras Ramos et al., 2014, in preparation, right panel):
large circles are the FLAMES targets. The superimposed black curve are the theoretical 
ZAHB models used to infer the atmospheric parameters.}
\label{cmd}
\end{figure*}

In this work  we analyzed a set of high-resolution spectra 
acquired with the multi-object spectrograph FLAMES in the MEDUSA/GIRAFFE mode
at the Very Large 
Telescope of the European Southern Observatory . 
The spectra are part of a dataset secured within a project aimed
at studying the general properties of blue straggler stars 
\citep{ferraro06,ferraro09a,ferraro12,lovisi12,lovisi13}. 
The employed GIRAFFE grating is HR5A (4340-4587 $\mathring{A}$, 
with a spectral resolution of $\sim$18000), suitable to sample
the He~I line at 4471.5 $\mathring{A}$. 
Spectra have been reduced with the standard ESO FLAMES pipeline. 
Six exposures of 45 min each have been secured in each cluster.
The SNR per pixel of the spectra around the He line ranges from $\sim$60 up to $\sim$130 
for M30, and from $\sim$75 up to $\sim$220 for NGC6397.
Radial velocity, atmospheric parameters and projected rotational velocity ($v_{e}$sini)
of each target have been derived and discussed in \citet{lovisi12,lovisi13} 
and we refer the reader to those papers for a detailed description.
Excluding stars with too noisy spectra and/or too low temperatures (for which 
the He~I line is not detectable), 
we are finally able to measure the He~I line in 24 stars of M30 and in 35 of NGC6397.
Fig.~\ref{cmd} shows the position of the targets in the 
color-magnitude diagrams of the two clusters (large circles) .
Table 1 lists their coordinates and atmospheric parameters.

\section{Chemical analysis}
\label{chem}

Stellar atmospheric parameters have been derived by \citet{lovisi12} 
and \citet{lovisi13} from the photometry. 
We recall the main information about the atmospheric parameters determination. 
$T_{eff}$ and logg have been derived by projecting the position of each 
star in the (V, V-I) plane on the best-fit theoretical Zero-Age Horizontal Branch (ZAHB) model. 
For NGC6397 the used ZAHB model is from the BaSTI dataset \citep{pietr06}, while for M30 
the ZAHB model is from the Pisa Evolutionary Library dataset \citep{cariulo}. For the latter, 
the choice of a different database of theoretical models is done for consistency with 
the analysis by \citet{ferraro09a}. However, we checked the consistency between the 
two sets of models: ZAHB models of the two databases chosen with the same metallicity
well overlap each other both in the observative and theoretical plane. The adoption of 
a dataset instead of another one leads to negligible changes in the atmospheric parameters, 
typically smaller than 30-40 K and 0.05 in $T_{eff}$ and logg, respectively 
(note that \citet{marino13} found a good agreement by using BasTI and PGPUC \citep{valcarce12} ZAHB models).
The used ZAHB models are shown in Fig.~\ref{cmd}.

\begin{figure}
\epsscale{0.9}  
\plotone{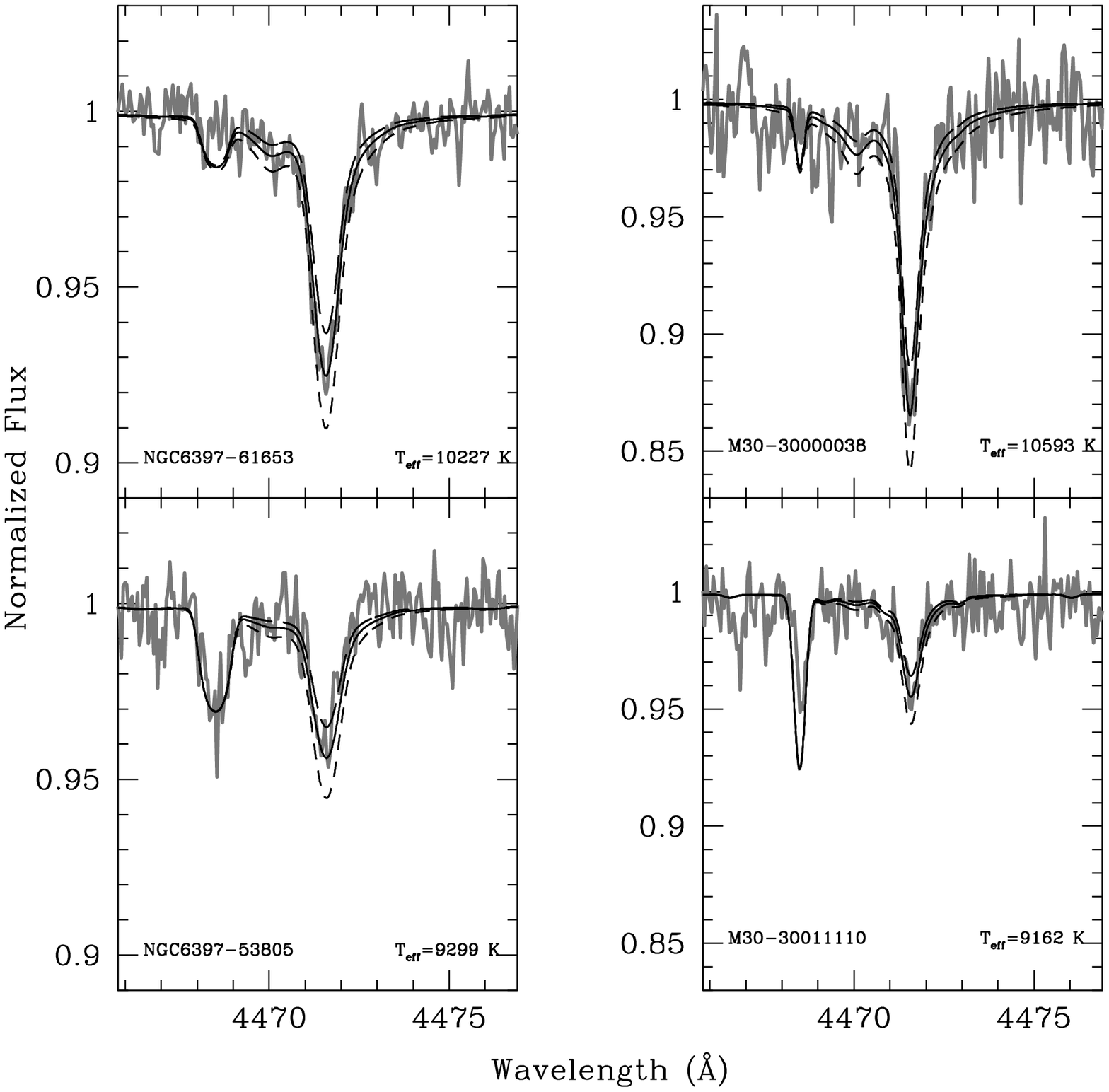}
\caption{Spectral region around the He line for the stars \#61653 and \#53805 in NGC6397 
and for the stars \#30000038 and \$30011110 in M30 (thick grey line), 
with overplotted synthetic spectra (black thin lines) calculated with the best-fit 
Y abundance and $\delta$Y=$\pm$0.1.}
\label{spec}
\end{figure}

The He abundance has been obtained for each target by fitting the observed He~I line 
at 4471.5 $\mathring{A}$ with a grid of synthetic spectra,  
calculated with the appropriate atmospheric parameters and varying only the He abundance.
The use of spectral synthesis (instead of the simple measure of the line equivalent width) 
is mandatory in the analysis of this line, to properly account for its relevant 
Stark broadening and to include the forbidden component at 4470 $\mathring{A}$ 
\citep[see e.g. ][]{mihalas74}.
Fig.~\ref{spec} shows the spectral region around the He line for one of 
the hottest and one of the coldest target stars in both clusters, 
with overplotted three synthetic spectra computed with the best-fit Y abundance and 
$\delta$Y=$\pm$0.1.

Synthetic spectra have been computed with the code SYNTHE \citep{sbordone}
adopting the line list provided by F. Castelli in her website. 
They have been convolved with a Gaussian profile 
in order to properly reproduce the spectral resolution of the 
HR5 grating and with a rotational profile in order to include 
the projected rotational velocities derived by \citet{lovisi12,lovisi13}. 

In order to properly take into account the contribution of the H and He abundances 
to the opacity, 
we calculated 
all the model atmospheres 
with the last version of the code 
ATLAS12\footnote{http://wwwuser.oat.ts.astro.it/castelli/sources/atlas12.html}
\citep{castelli05}. At variance with the widely 
used ATLAS9 code (that adopts pre-tabulated opacities calculated for specific chemical mixtures, 
in particular with standard He mass fraction Y=~0.245), 
ATLAS12 employs the opacity sampling method \citep{peytremann74}
and allows one to calculate model atmospheres with arbitrary chemical composition.
All the model atmospheres have been computed under the assumption of Local 
Thermodynamical Equilibrium (LTE) and one-dimensional, plane-parallel geometry.
We checked the impact of the use of ATLAS9 and ATLAS12 models on the derived He abundance. 
For Y around the standard value (Y$\sim$0.25) the two models provide the same result, 
while for He-enhanced stars (at least up to Y$\sim$0.3), the adoption of ATLAS9 models 
under-estimates Y of about 0.02. On the other hand, for stars with surface He mass fraction 
of $\sim$0.10, analysis based on the standard ATLAS9 models overestimate Y by $\sim$0.05.

Despite the analyzed He transition can suffer for departures from LTE conditions
(relevant for B-type stars), this effect is negligible considering 
the atmospheric parameters and the metallicities of our targets 
(P. Bonifacio, private communication).

The total uncertainty for each star is derived by adding in quadrature 
the uncertainty in the fitting procedure and that arising from the 
adopted parameters.
The uncertainty in the fitting procedure has been estimated 
by using MonteCarlo simulations. For observed spectra, the uncertainty associated 
to a $\chi^2$-minimization cannot be estimated by using the $\chi^2$ theorems, that assume that all 
the pixels are not correlated each other \citep[see e.g. the discussions in][]{cayrel99,caffau05}. 
In fact, in observed spectra the adjacent pixels cannot be considered as independent each other because of 
the re-binning procedure during the wavelength calibration.
For each star, we computed a set of 1000 synthetic spectra, calculated 
with the appropriate atmospheric parameters, rotational velocities and the best-fit He abundance.
Each MonteCarlo spectrum has been obtained by rebinning the best-fit synthetic spectrum to the same pixel-size 
of the GIRAFFE spectra (0.05 $\mathring{A}$/pixel)
and then by injecting Poissonian noise, in order to reproduce the SNR of each 
star around the He line. Thus, this set of synthetic spectra is equivalent to the real one but 
with the He abundances known a priori. This method allows to take into account simultaneously the 
main sources of uncertainty in the line fitting, namely the finite size of the pixels, the SNR and the 
continuum estimate.
The same analysis performed for the observed spectra has been done for the 
synthetic ones and
the dispersion of the derived Y abundance distribution has been assumed as 1$\sigma$ 
uncertainty in the fitting procedure. These uncertainties depend on the 
injected SNR, but also on the line strength (thus the temperature and the He abundance) 
and the rotational velocity.
Typical errors in He mass fraction range from 0.01 up 0.05. 

Because we are interested in possible star-to-star variations of the He content, we 
estimated the internal uncertainties due to the atmospheric parameters. 
The total error obtained by adding in quadrature the uncertainties due to the individual 
atmospheric parameters is an upper limit of the internal error, because it does not take 
into account the covariance terms occurring among the parameters. 
In order to take into account  the effect of the projection process on the 
derived $T_{eff}$ and log~g, we adopt the following procedure:
we re-projected each target on the best-fit ZAHB by including its photometric uncertainty,  
re-determining simultaneously $T_{eff}$ and log~g, in order to include the correlation between 
the two parameters. With this method we derive variations in $T_{eff}$  between 
$\sim$70 and $\sim$150 K, with corresponding variations in gravity of the order of 
0.02. These relatively small uncertainties in $T_{eff}$ and log~g are  
essentially due to the high internal accuracy of the adopted photometric catalogs, with 
typical photometric uncertainties of $\sigma$(V-I)$\sim$0.01-0.02 mag, 
obtained by the average of several independent measures \citep[see e.g.][]{ferraro09a}.
Note also that these uncertainties do not represent the total error budget in the adopted parameters, 
but only the internal star-to-star uncertainty related to the adopted procedure in the parameter derivation.

Only to provide the general variation of Y due to this procedure, 
an uncertainty of $\pm$100 K in $T_{eff}$ 
(coupled with the corresponding variation in gravity of $\pm$0.02) provides a variation in Y of $\pm$0.01 for the hottest
stars ($\sim$11000 K) 
and of $\pm$0.02 for the coldest targets ($\sim$9000 K), whereas the impact of 
microturbulent velocity is totally negligible.
The error in $v_{e}$sini (typically 2-3 km/s) provides a contribution at a level of 
less than 0.005. 

Note that the 4471 $\mathring{A}$ He line used in this work is slightly less sensitive to the 
adopted atmospheric parameters with respect to the line at 5875 $\mathring{A}$, adopted 
in the other papers where the He abundance in GC stars is derived.




 Finally, systematic effects can be due to the choice of the ZAHB model.
As extensively discussed by \citet{marino13}, a possible source of systematic errors is the He abundance of the 
used ZAHB. The He abundance of our targets is not known a priori, thus we derived the atmospheric parameters 
adopting ZAHB models computed with standard Y. The adoption of a Y-enhanced ZAHB leads 
to a decrease of gravity by $\sim$0.1-0.15, with a negligible impact on the temperature. 
Note that a systematic decrease of 0.1 in log~g (keeping $T_{eff}$ fixed) implies an increase of the derived Y smaller than 0.02/0.03. 
As discussed in Section \ref{content}, the adoption of the standard He content for the used ZAHB models is reasonable 
in light of the derived He content of our targets, thus we do not need to re-derive the atmospheric 
parameters by using ZAHB models computed with higher Y. 
A similar effect is obtained if we consider that the stars leaving the ZAHB locus will be 
more luminous and with a lower gravity (but basically the same temperature) with respect 
to the ZAHB position.


\section{The He content of M30 and NGC6397}
\label{content}

\begin{figure}
\epsscale{0.9}  
\plotone{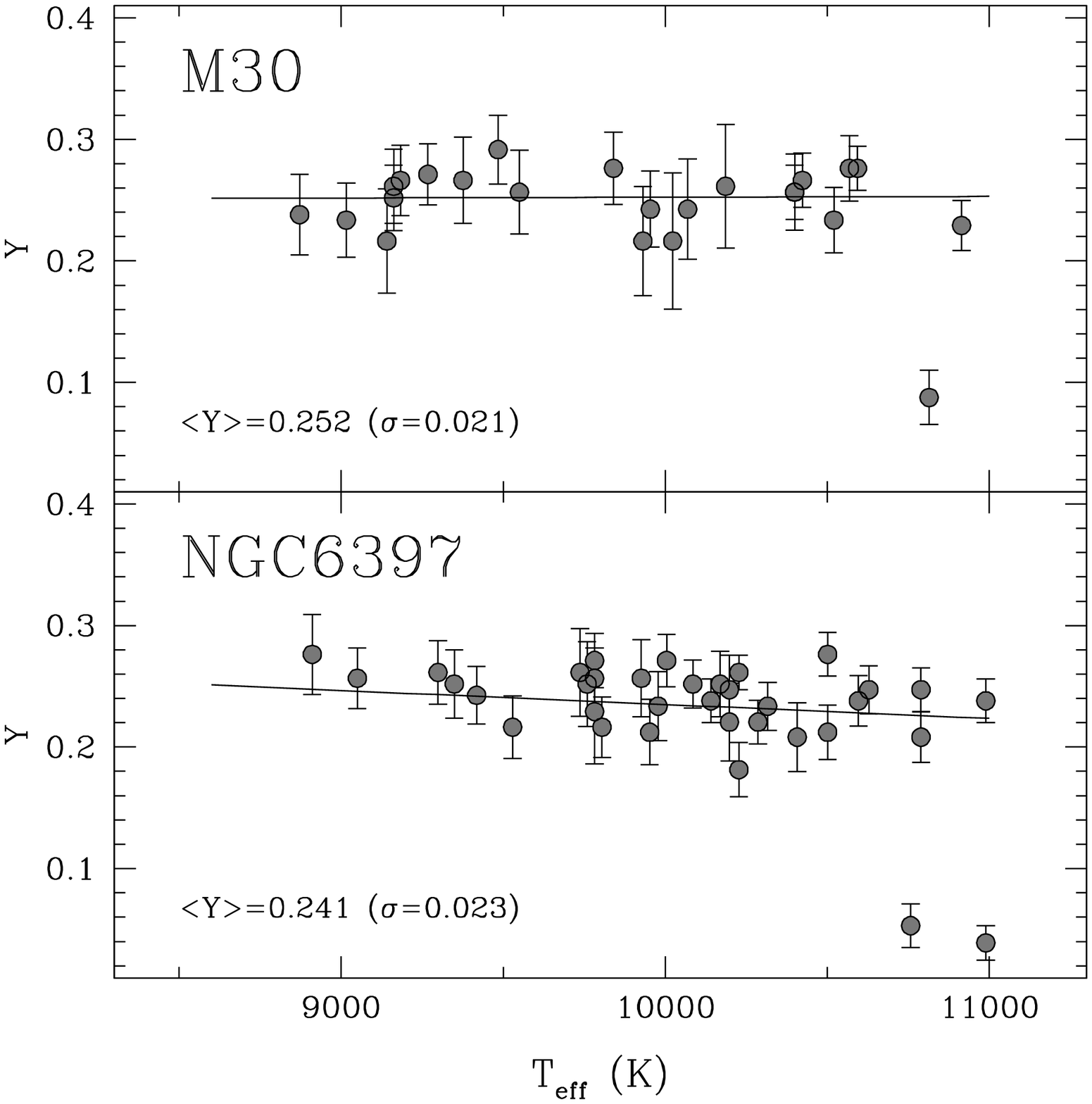}
\caption{
Behavior of the He mass fraction Y as a function 
of the temperature for the HB stars of M30 (upper panel) and NGC6397 (lower panel). 
Solid lines are the linear fits calculated excluding the stars with evidence of 
radiative levitation.}
\label{res1}
\end{figure}

\begin{figure}
\plotone{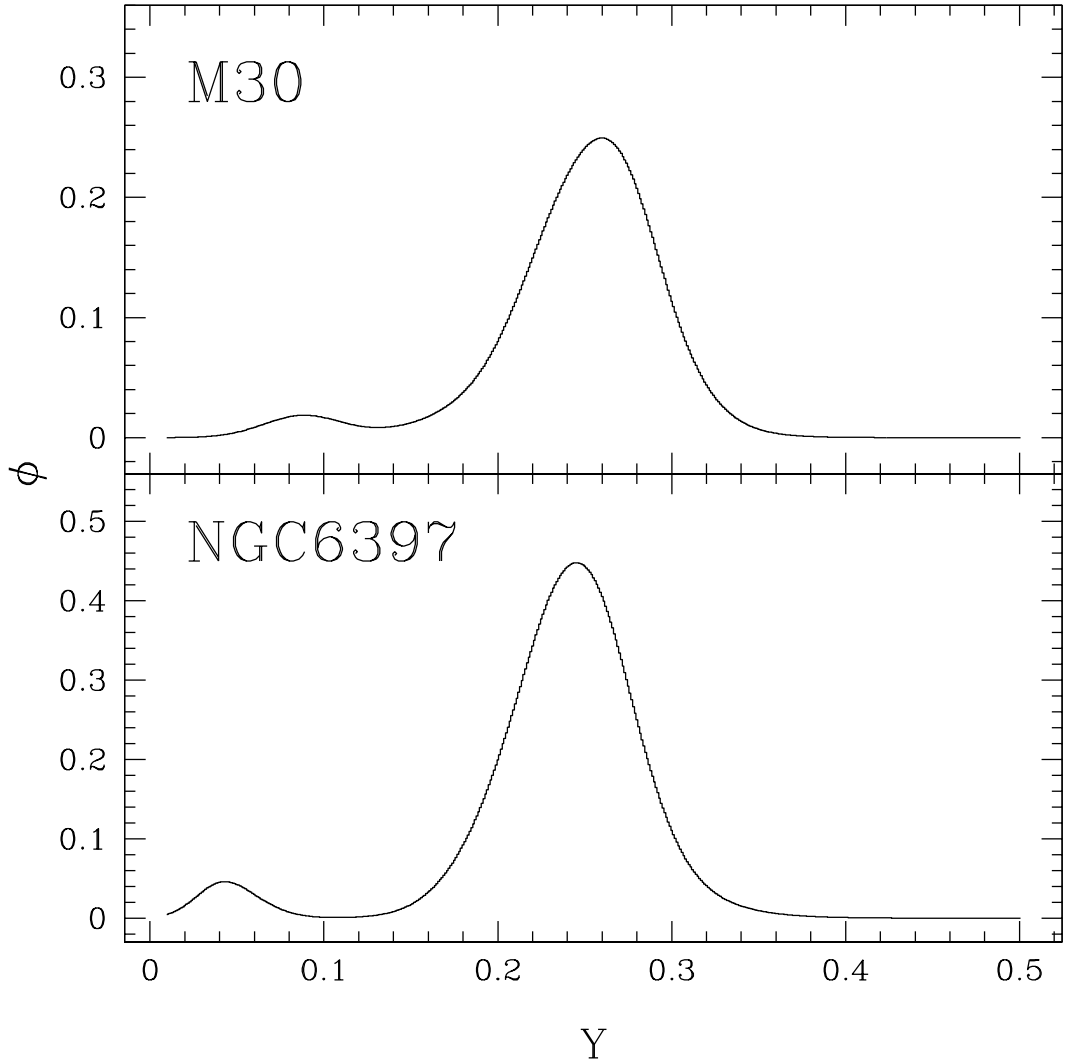}
\caption{Generalized histograms for the He mass fraction 
of the HB stars in M30 (upper panel) and in NGC6397 (lower panel).
}
\label{res2}
\end{figure}

Table 1 lists the derived He mass fraction of the targets and their total uncertainty.
Fig.~\ref{res1} shows the behavior of Y as a function of the temperature for the stars 
of M30 (upper panel) and NGC6397 (lower panel), 
while Fig.~\ref{res2} shows the Y distributions in the two samples of stars represented as 
generalized histograms 
\citep[a representation that removes the effect due to the choice of the starting point 
and of the bin size, and takes into account the individual uncertainty of each star; see][]{laird}.
In both GCs, all the stars have Y around $\sim$0.24-0.25, with the exception 
of one star in M30 and two stars in NGC6397, that show very low (Y$<$0.1) He abundance.
The three stars with low He content also show iron abundances higher than
that of the parent cluster \citep{lovisi12,lovisi13}.
This behavior is commonly observed in HB stars hotter than the
{\it Grundahl Jump}  \citep{behr1999, behr2000, hubrig09, gratton12b} 
and it is predicted by theoretical models \citep{michaud83,quievy}
as an effect of radiative levitation (responsible for the metal enhancement) 
and gravitational settling (responsible for the He depletion). 
We note that some stars with similar $T_{eff}$ to those of the He-poor stars, but with 
normal Y, are detected. This difference can be due to the fact that we observe 
the region close to the {\sl Grundahl Jump} and not all the stars have still undergone 
the diffusion processes. 
Interestingly enough, one star in NGC2808 with a temperature higher than that of the {\sl Grundahl Jump} 
does not show any evidence of He depletion \citep{marino13}.

Excluding the three Fe-rich and He-poor stars, 
we find average He mass fractions of Y=0.252$\pm$0.005 ($\sigma$=0.021) for M30 and 
Y=0.241$\pm$0.004 ($\sigma$=0.023) for NGC6397. 
{\it It is worth to notice that these are not only the first determinations of Y 
for M30 and NGC6397, but they are also the first ones for GCs with [Fe/H]$<$--2.0 dex}
\footnote{\citet{behr2000} and \citet{behr03} identified one HB star in M92 and one HB star 
in M15 (both [Fe/H]$<$--2.0 dex), not affected by levitation and gravitational settling 
effects. However, 
their huge uncertainties ($\sim$0.3-0.4 dex) do not allow to firmly establish the real He content 
of these GCs.}.

\begin{figure}
\plotone{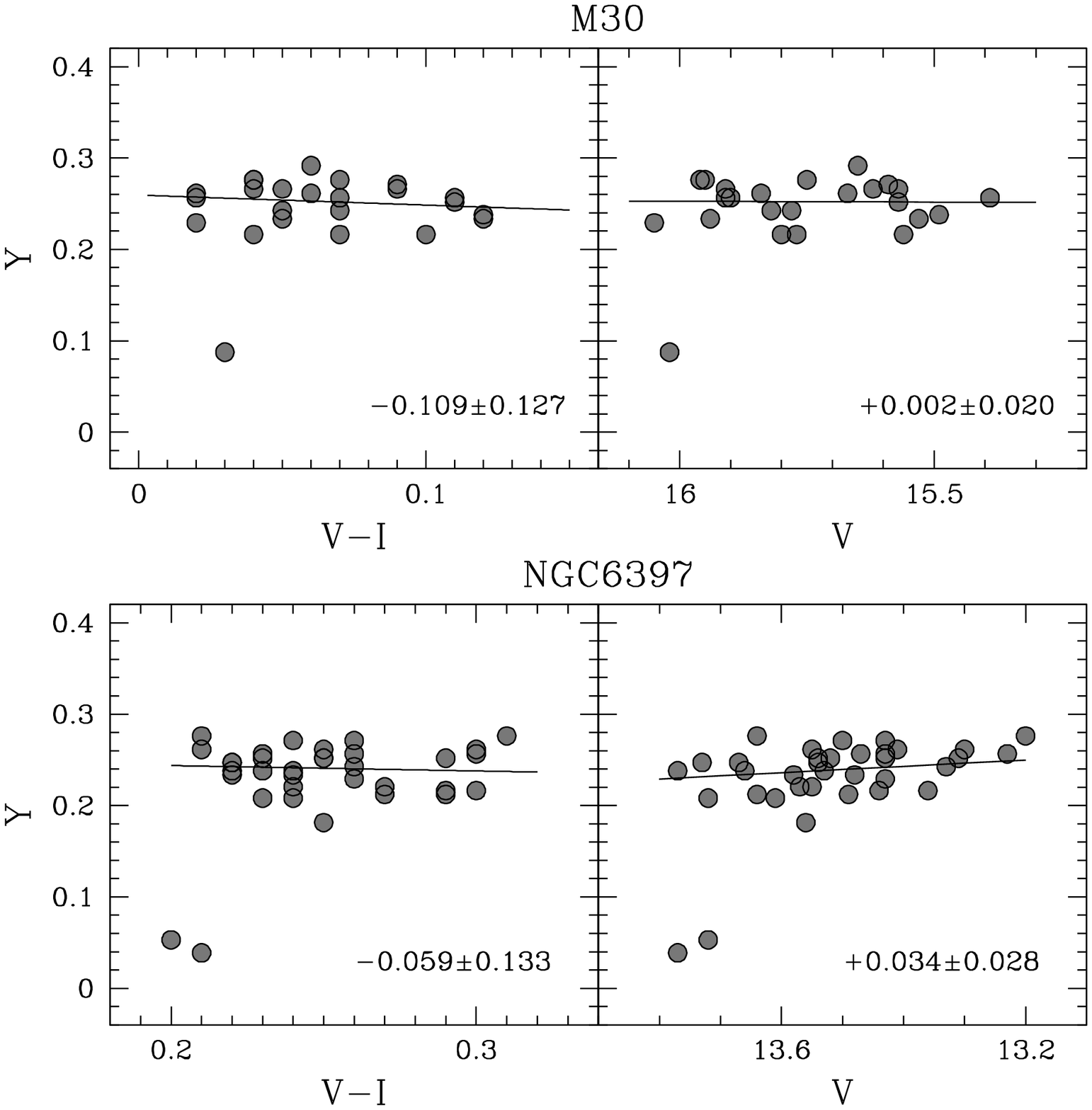}
\caption{
Behaviour of the He mass fraction Y as a function of the V-I color (left panel) and 
of the V-band magnitude (right panel) for M30 (upper panels) and for NGC6397 (lower panels). 
Solid lines are the best-fit linear fits (the corresponding slopes and uncertainties are labelled).}
\label{pht}
\end{figure}


According to the theoretical models of \citet{pietr06}, the surface 
He mass fraction for a star with 0.8 ${\rm M}_{\odot}$, Z=~0.0003 (corresponding to [Fe/H]=--2.1)
and $\alpha$-enhanced chemical mixture, increases by only 0.01 with respect to the 
initial value, after the First Dredge-Up episode.
Therefore, the derived He abundances of HB stars in M30 and NGC6397 are totally 
compatible with the expectations for low-mass evolved stars formed with a primordial 
He abundance \citep[$Y_{P}$=~0.254, ][]{izotov}.


No trend between the He abundances and the corresponding (V-I) color and V-band magnitude 
is detected for the stars with no evidence of radiative levitation . Fig.~\ref{pht} shows 
the behaviour of Y as a function of (V-I) and V. 
The best-fit linear fits are calculated with the routine {\tt fitexy} 
by \citet{press} to take into account the uncertainties in both 
the quantities, whereas the corresponding 
uncertainties in the slope are calculated with the Jackknife bootstrapping technique. In a similar way, no evident 
trend between Y and $T_{eff}$ is recognized:  Fig.~\ref{res1} shows the linear fits, providing 
slopes of 6.4$\cdot10^{-7}\pm$0.005 and -1.15$\cdot10^{-5}\pm$0.008.

The observed Y values among the stars of each target GC are compatible within the 
uncertainties. Thus, we can conclude that the two GCs are not strongly enriched in He, 
displaying a substantial He uniformity: only small (if any) Y variations could be 
present in their stellar content.
This result agrees with the analysis of NGC6397 by \citet{dicri}, 
based on the width of the observed main sequence (MS), that predicts
a maximum internal variation of $\sim$0.02 in the Y distribution of 
the cluster MS stars. 
Further results by \citet{milone12} revealed the presence of a double MS
in the color-magnitude diagram of NGC6397.
This can be reproduced with a population (accounting for 30\% of the total 
cluster population) 
having normal Y and another one with a mild He-enhancement of about 0.01. 
Again, this is fully consistent with our results.
Concerning M30, no study so far has revealed splitting or anomalous 
broadening of the MS, suggesting a small or null intrinsic dispersion 
in the He content of this cluster, in agreement with our findings.

Finally, the uniform He content that we find in the HB stars of M30 and NGC6397 well 
agrees with theoretical models that predict only a mild He enhancement for clusters 
with HB morphologies similar to that of our targets (covering a narrow extension in color, thus 
in $T_{eff}$; see Fig.~\ref{cmd}), at odds with clusters with very extended blue tails 
for which high He enhancements (Y$\ge$0.30) are predicted \citep[see e.g. Fig.~10 in][]{dalex13}.


\section{He abundance and self-enrichment process}

He enrichment in GC sub-populations is expected in light 
of the self-enrichment processes, thought to occur during the early stages 
(within $\sim$100 Myr) of GC history.
All the GCs studied so far, both in the Milky Way \citep{c09} and in other 
galaxies of the Local Group \citep{letarte,m09}, display 
well-established chemical patterns, with homogeneous iron-peak element abundances
and with anticorrelations between C and N, between O and Na, and 
(for some clusters) between Mg and Al. 
The only exceptions are a bunch of peculiar GC-like systems with 
an intrinsic dispersion in their iron content (with broad and/or multimodal [Fe/H] distributions), namely 
Terzan 5 \citep{ferraro09b}, Omega Centauri 
\citep[see e.g.][]{johnson,pancino11a,pancino11b}, 
M22 \citep{marino09,marino11b} and M54 \citep{bell,carretta_m54}
\footnote{Note that other GCs are suspected to have small iron dispersions, 
namely NGC1851 \citep{carretta_1851}, NGC5824 \citep{saviane12} and NGC3201 
\citep{simmerer13,munoz13}, but there is no general consensus about them.}.
The chemical patterns involving light elements and observed in GCs 
are commonly interpreted as the signature of material 
processed through the high temperature extension of the proton-capture reactions 
(like NeNa and MgAl cycles).

Intermediate-mass AGB stars \citep{dercole} and fast-rotating, massive stars \citep{decressin},
both able to ignite the complete CNO-cycle, have been proposed as main polluters.
Whichever the true nature of the polluters is, 
new cluster stars, formed from pristine gas diluted with material 
processed in the stellar interiors, are expected to be also enriched in He, with 
a level of He enrichment varying from cluster to cluster, 
from very small values ($\le$0.02), as in the case of NGC6397 \citep{dicri, milone12}, 
up to extreme He contents (Y$\sim$0.4), as those proposed to explain the 
complex MS and/or HB morphologies observed 
in $\omega$ Centauri \citep{piotto05}, NGC2808 \citep{dantona05,piotto07,dalex11} and 
NGC2419 \citep{dicri11}.

A first, indirect hint of Y-[O/Fe] anti-correlation has been provided 
by \citet{villanova09} and \citet{villanova12}, who analyse 
red HB stars of NGC6752 and blue HB stars of M4, respectively. The HB stars 
in NGC6752 show enhanced [O/Fe] ratios and Y compatible with the 
cosmological value, while the stars along the blue portion of the HB in M4 
have enhanced values of Y (by 0.04-0.05) and [O/Fe] ratios compatible with 
the second generation stars of the cluster.
Even if performed on two different clusters, these results by \citet{villanova09} and 
\citet{villanova12} suggest that the blue part of the HB is mainly populated by stars formed 
from gas enriched in He and, generally speaking, by the products of the high temperature 
proton-capture reactions. Analysis based on other elements and not involving directly the measure of the 
He abundance, have confirmed the connection between the position of the HB stars and their 
chemical composition \citep[see for instance][for the cases of M4 and NGC1851, respectively]{marino11,gratton12b}.

\begin{figure}
\epsscale{0.9}  
\plotone{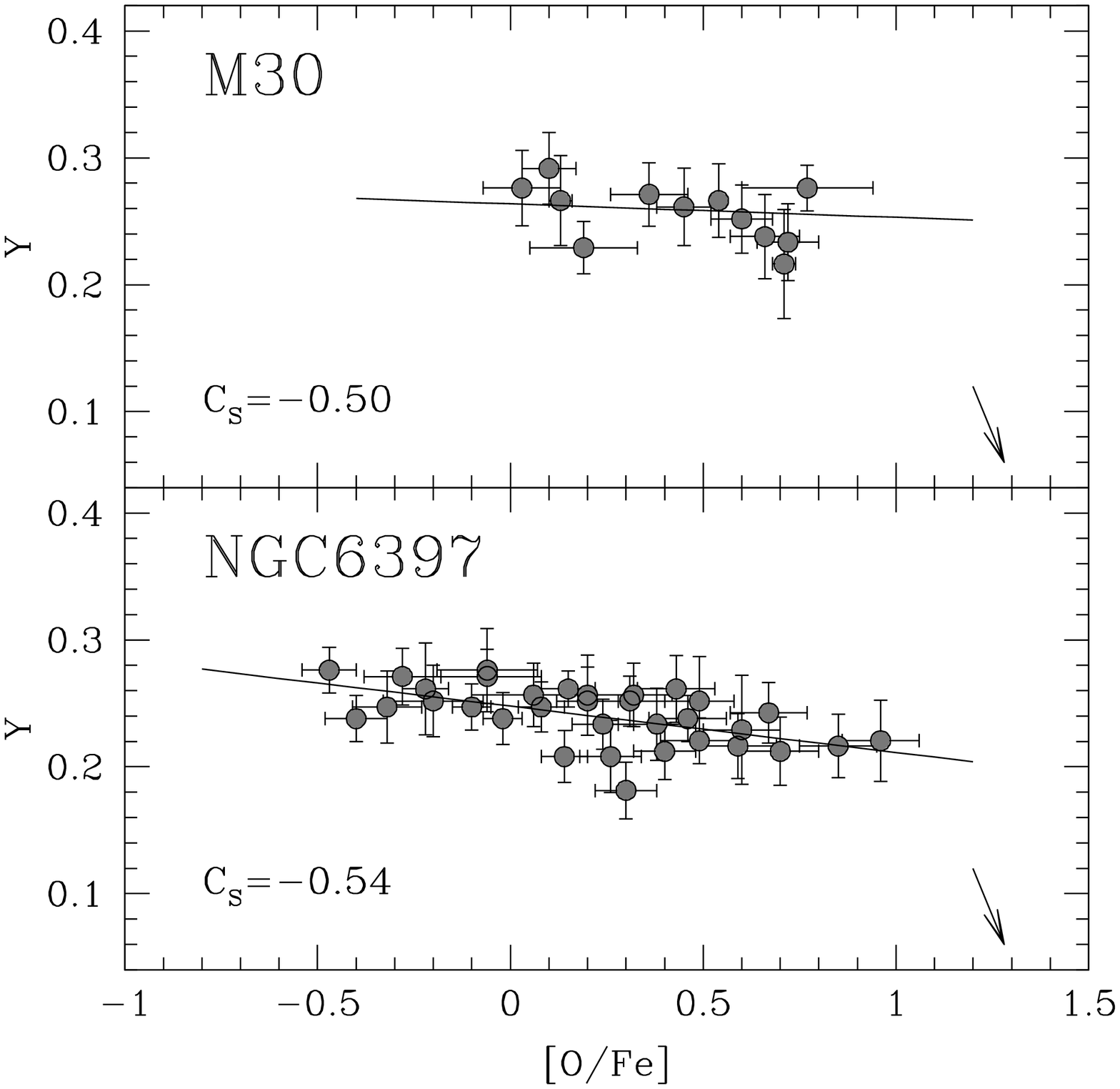}
\caption{Behavior of the Y abundance as a function of 
[O/Fe] for the stars in M30 (upper panel) and NGC6397 (lower panel).
The arrows show the effects of a change in 
$T_{eff}$ and logg. Labelled are the values of the Spearman correlation 
coefficient. Solid lines are the best-fit straight lines.
}
\label{hox}
\end{figure}

We can use our dataset to probe the existence of any Y-[O/Fe] correlations 
in the two surveyed clusters. Indeed \citet{lovisi12} and \citet{lovisi13} 
measured non-LTE [O/Fe] abundances 
for several HB stars of the two target clusters from the oxygen triplet 
at $\sim$7770 $\mathring{A}$. 
Fig.~\ref{hox} shows the behavior of Y as a function of [O/Fe] 
(excluding the Y-poor stars where the radiative levitation and gravitational 
settling have modified the surface abundances). 

Abundances of both [O/Fe] and Y are available for only 12 stars of M30.
No correlation between the two abundances is detectable 
(upper panel in Fig.~\ref{hox}): 
a straight line fit, 
performed with the routine {\tt fitexy} by \citet{press} 
provides a slope of -0.011$\pm$0.057 
\citep[where the uncertainty is computed with a Jackknife bootstrapping technique; see][]{lupton}.
The small probability of correlation is confirmed also by
the Spearman rank correlation coefficient ($C_S$=--0.50), leading 
to a probability of only 90\% that the two abundances are correlated.

On the other hand, the sample of 33 stars of NGC6397 for which both O and Y 
are available displays a mild Y-[O/Fe] anti-correlation 
(lower panel in Fig.~\ref{hox}). 
A linear fit provides a slope of -0.036$\pm$0.010, corresponding to a 3.6$\sigma$ detection. 
The Spearman rank correlation coefficient is $C_S$=--0.54, providing a 
probability higher than 99.9\% of an anti-correlation between the two abundances.
The same result is confirmed also by a non-parametric Kendall-$\tau$ test
\footnote{In a similar way, \citet{monaco} recognized a very mild anti-correlation between Na and Li 
abundances among the dwarf stars of M4 (and justified in the framework of the multiple 
populations in GCs). Even if their abundance distributions do not show evidences of 
intrinsic scatter (in light of the estimated uncertainties), both parametric and 
non-parametric rank correlation test highlight an anti-correlation between the two abundances.}.

An interesting difference between the two clusters is their [O/Fe] distributions, 
being that of NGC6397 larger than that of M30 and including a component with [O/Fe]$<$0. 
Previous determinations of the O abundance in NGC6397 provide a small range of [O/Fe], 
with no evidence so far of O-poor stars.
Despite its proximity, the number of stars in NGC6397 in which the O abundance has been 
measured is very small and most of the analysis available so far are based on the 
forbidden O line at 6300 $\mathring{A}$.
\citet{castilho} provided [O/Fe] for 2 (out of 16) giants, 
finding for both the stars [O/Fe]=+0.15 dex. 
\citet{c09,c09b} properly measured O in 12 giants observed with UVES (reaching [O/Fe]=+0.11 dex) 
and provided upper limits for other 7 giants, while for most of the stars of their GIRAFFE survey 
no measures at all are provided, because of the low SNR and the radial velocity 
of the cluster (RV$\sim$20 km/s) that leads to an overlap between the forbidden O line with the sky O 
emission line. Recently, \citet{lind11} derived O abundances for 16 giant stars, finding 
a very small variation of O among their stars, from [O/Fe]=+0.41 up +0.77 dex.  
Only \citet{gratton01} measured the oxygen triplet at 7770 A for 7 dwarf/subgiants 
(and an upper limit) finding a range between +0.08 and +0.48 dex.

However, we suggest a possible bias in the measure of the O distribution of NGC6397 from giant stars.
The derivation of the precise [O/Fe] abundances range in the giant stars of metal-poor globular clusters
can be quite complex, because the only available oxygen line is the forbidden one that is very weak at low 
metallicity. Moreover, the almost zero radial velocity makes 
impossible to properly detect the O line (in the case of M~30 this effect does not occur because 
of its radial velocity, ~-185 km/s, prevents any blending with the sky emission line). 
We conclude that the giant stars are not the best sample to properly study the O abundance 
(and in particular to identify the most O-poor stars) in NGC6397. 
If a [O/Fe] sub-solar component does exist among the star of NGC6397, it cannot be detected from the 
analysis of its giant stars.
On the other hand, the O triplet at 7770 A is well detectable and strong among HB stars, providing 
a more robust diagnostic. 
Also, we note the very good match between our [O/Fe] distribution and that 
by \citet{c09b} for M30, where the very low radial velocity of this cluster prevents 
any spurious blending between the forbidden O line and the emission O sky line.

We checked whether the impact of the atmospheric parameters uncertainties 
is able to introduce a spurious anti-correlation between the two abundances. 
In fact, the increase of $T_{eff}$ (coupled with the corresponding 
increase of log~g) leads to an increase of [O/Fe] and a decrease of Y.
However, the slope is significantly steeper (-0.75) than that observed for the stars in NGC6397 
(see the arrows in Fig.~\ref{hox}, showing the effects of a change in $T_{eff}$ and log~g
by --200 K and --0.04, respectively). 
This slope remains the same for stars with different atmospheric parameters and 
with difference O abundances.
Thus, we can rule out that the observed anti-correlation is an artifact 
of the uncertainty of the atmospheric parameters. 
Also, we checked that no correlation does exist between the abundances and $v_{e}$sini; 
note that the internal uncertainties in $v_{e}$sini are not able to introduce a spurious anticorrelation 
between the abundances.

As a sanity check, we roughly divided the Y abundances of NGC6397 in two samples, 
corresponding to [O/Fe] lower and higher than the solar value, finding 
$<$Y$>$=~0.258$\pm$0.005 ($\sigma$=~0.015) and 
$<$Y$>$=~0.233$\pm$0.005 ($\sigma$=~0.022), respectively. 
This small difference \citep[formally compatible with the results by][]{dicri,milone12} 
corresponds to a 3.5$\sigma$ detection.
A Kolmogorov-Smirnov 
test provides a $\sim$1\% probability that the Y abundances of the stars with sub-solar 
[O/Fe] abundances are extracted from the same population as the stars with [O/Fe]$>$0.0.

\section{Summary}

We have analysed the He mass fraction Y for a sample of 24 and 35 HB stars in M30 and NGC6397, 
respectively. The main results are: {\sl (i)} both clusters have an average He content 
compatible with the primordial He abundance ($<$Y$>$=0.252$\pm$0.003 for M30 and 
$<$Y$>$=0.241$\pm$0.004 for NGC6397) and they are not strongly enriched in He; 
{\sl (ii)} a weak (but statistically significant) anticorrelation between Y and [O/Fe] 
among the HB stars of NGC6397 does exist (but it is not detected in M30).

We suggest that the O-poor, He-rich stars found in the HB of NGC6397 belong 
to the second stellar generation of the cluster. 
Unfortunately Na abundances are not available for these stars.
In principle, Y-[O/Fe] anti-correlation is expected in all the GCs displaying 
the chemical signatures of the self-enrichment processes, even if its very small 
slope makes its detection very hard.
The lack of Y-[O/Fe] anti-correlation for the stars in M30 can be due to 
several causes, mainly the size of our sample (three times smaller than that secured for NGC6397) 
and the SNR of the spectra (lower than that of the spectra of NGC6397). Also, we cannot 
rule out that M30 has undergone a self-enrichment process less efficient with respect 
to NGC6397, as suggested by their different [O/Fe] distributions (in fact 
M30 shows a lack of stars with [O/Fe]$<$0, instead detected among the stars of NGC6397). 
Thus, the internal variation of the He content in the stellar population of M30 
could be smaller than 0.01.

The Y-[O/Fe] anti-correlation observed in NGC6397 seems to confirm the 
theoretical expectations that the GC stars born after the first burst of star formation 
are both depleted in O and (mildly) enriched in He, demonstrating
that the stars usually labelled as {\it second generation stars} 
show the signatures of hot-temperature proton-capture processes, with a 
simultaneous O-depletion and a weak He enrichment.

\acknowledgements  
The authors warmly thanks the anonymous referee for his/her helpful comments that improved 
the quality of the paper. 
Also, we are grateful to P. Bonifacio for his useful suggestions about the NLTE corrections.
This research is part of the project COSMIC-LAB funded by the European Research Council 
(under contract ERC-2010-AdG-267675).

\newpage

\begin{deluxetable}{ccccccc}
\tablecolumns{7} 
\tablewidth{0pc}  
\tablecaption{Stellar parameters and He abundances.}
\tablehead{
\colhead{ID} &      Ra  &   Dec     & \colhead{$T_{eff}$} & \colhead{log~g} & \colhead{Y}  & \colhead{$\sigma_{Y}$} \\
             &  (J2000) & (J2000)   &        (K)	    &		      & 	     &  		     }
\startdata 
\hline
M30 &  &  &  &  &  &  \\
\hline   
10201925  &  325.0989565 & -23.1613142  &   10914  &	 3.7  &  0.229  &  0.021  \\
10202614  &  325.0891261 & -23.1711663  &   10069  &	 3.6  &  0.243  &  0.041  \\
10203922  &  325.0904138 & -23.1512884  &   10186  &	 3.6  &  0.261  &  0.051  \\
10301333  &  325.1135774 & -23.1751181  &   10023  &	 3.6  &  0.216  &  0.056   \\
10301793  &  325.1201773 & -23.1736029  &    9376  &	 3.4  &  0.266  &  0.036  \\
10400762  &  325.0974077 & -23.1873751  &    9226  &	 3.4  &  0.208  &  0.036  \\
10401890  &  325.1056998 & -23.1838967  &    9931  &	 3.6  &  0.216  &  0.045  \\
\hline		 
\enddata 	 
\tablecomments{Identification numbers, coordinates, temperatures, gravities \citep{lovisi12,lovisi13}, 
He mass fractions and uncertainties for the observed stars. A complete version of the table 
is available in electronic form.\\}
\end{deluxetable}

\end{document}